\documentclass{csmagclass}													

\begin{document}		
\headings																															 %

\title{{ Comparative Study of a Critical Behavior of a Coupled Spin-Electron Model on a Doubly Decorated Square Lattice in the Canonical and Grand-Canonical Ensemble}}
\author[1]{{ H. \v{C}en\v{c}arikov\'a\thanks{Corresponding author: hcencar@saske.sk}}}
\author[1]{{ N. Toma\v{s}ovi\v{c}ov\'a}}
\affil[1]{{Institute of Experimental Physics, Slovak Academy of Sciences, Watsonova 47, 040 01 Ko\v{s}ice, Slovakia}}

\maketitle

\begin{Abs}
The critical behavior of a hybrid spin-electron model with localized Ising spins  placed on nodal sites and mobile electrons delocalized over bonds between two nodal lattice sites is analyzed by the use of a generalized decoration-iteration transformation. Our attention is primarily concentrated on a rigorous analysis of a critical temperature in canonical and grand-canonical statistical ensemble at two particular electron concentrations, corresponding to a quarter ($\rho\!=\!1$) and a half ($\rho\!=\!2$) filled case. It is found that the critical temperature of the investigated spin-electron system in the canonical and grand-canonical ensemble may be remarkably different and is very sensitive to the competition among the model parameters like the electron hopping amplitude ($t$), the Ising coupling between the localized spins ($J'$), the electrostatic potential ($V$) and the electron concentration ($\rho$). In addition, it is detected that the increasing electrostatic
  potential has a reduction effect upon the deviation between critical temperatures in both statistical ensembles.
\end{Abs}
\vspace*{-0.5cm}\section{Introduction}\vspace*{-0.2cm}
The study of the critical behavior is one of the most extensively studied problem in a condensed matter physics, with an aim to deeper understand processes realized in the vicinity of a critical point. The most effective treatment from the theoretical point of view is the application of traditional analytical approaches or state-of-the-art numerical methods on suitable lattice-statistical models. Based on the method as well as model complexity, the choice of the statistical ensemble may be crucial to resolve a defined mathematical problem. Although in most textbooks the equivalence between different ensembles in the thermodynamic limit is illustrated via the Van Hove theorem~\cite{Huang}, there exists a few studies which indicate on the ensemble inequality (see Ref.~\cite{Mehdi} and Refs.~5-17 therein).
In the present work, we will examine a relatively simple doubly decorated spin-electron model on a square lattice (DDSEM), which  was successfully used to elucidate  the origin of some unconventional phenomena in coupled spin-electron systems like a doping-induce crossover from ferro- to antiferromagnetism~\cite{Doria,Strecka}, an enhanced magnetoelectric effect~\cite{Cenci1}, metamagnetic transitions~\cite{Cenci3} or magnetic re-entrance~\cite{Cenci2}. Our main attention will be concentrated on the comparative study of a critical behavior in the canonical ensemble ($CE$) and grand-canonical one ($GCE$), with a motivation to examine if the choice of the statistical ensemble could have a significant impact on a critical temperature in the DDSEM, where the intrinsic features can be modified by an extrinsic factor like an electric field.
\vspace*{-0.25cm}
\section{Model and methods}
\vspace*{-0.25cm}
Let us consider an interacting spin-electron model on the square lattice (see Fig.~1 in Ref.~\cite{Cenci2020}), which consists of Ising spins localized at nodal lattice sites and mobile electrons delocalized over two decorating atoms between two nearest-neighbor Ising spins. Due to the local character of all assumed interactions, the total  Hamiltonian can be divided into the sum of 2N commuting bond Hamiltonians $\hat{\cal H}\!=\!\sum_{k=1}^{2N}\hat{\cal H}_k$, where
\vspace*{-0.3cm}
\begin{eqnarray}
\vspace*{-1cm} 
\hat{\cal H}_k\!\!\!\!&=&\!\!\!\!-t\sum_{\gamma=\uparrow,\downarrow}(\hat{c}^{\dagger}_{k_1,\gamma}\hat{c}_{k_2,\gamma}\!+\!
H.c.)\!-\!J'\hat{\sigma}^{z}_{k_1}\hat{\sigma}^{z}_{k_2}
\!-\!
\sum_{\alpha=1,2}\left[J\hat\sigma^{z}_{k_{\alpha}}(\hat{n}_{k_{\alpha},\uparrow}\!-\!\hat{n}_{k_{\alpha},\downarrow})\!-\!(\!-\!1)^{\alpha}V/\sqrt{2}\hat{n}_{k_{\alpha}}\right].
\label{eq1}
\end{eqnarray}
\vskip -0.3cm
Here $\hat{c}^{\dagger}_{k_{\alpha},\gamma}$ and $\hat{c}_{k_{\alpha},\gamma}$ $(\alpha\!=\!1,2; \gamma\!=\!\uparrow,\downarrow)$ denote the usual creation and annihilation fermionic operators, for which the respective number operators read $\hat{n}_{k_{\alpha}}\!=\!\sum_{\gamma}\hat{n}_{k_{\alpha},\gamma}$ and $\hat{n}_{k}\!=\!\sum_{\alpha}\hat{n}_{k_{\alpha}}$. In addition, $\hat{\sigma}^z_{k_{\alpha}}$ is the $z$ component of the Pauli operator with the eigenvalues $\sigma^z_{k_{\alpha}}\!=\!\pm1$.  The Hamiltonian~(\ref{eq1}) is formed by the kinetic energy of mobile electrons modulated by the electron hopping ($t$) and exchange interactions of Ising-type   between (i) the  nearest-neighbor Ising spins ($J'$) and (ii)  the Ising spins and their nearest-neighbor mobile electrons ($J$).  Finally, the  Hamiltonian also involves the electrostatic energy for a pair of mobile electrons  modulated through the electrostatic potential $V$ originating from the applied electric field acting along the crystallographic axis \textbf{[11]}. This spatial field orientation results in the identical influence of electric dimers on horizontal and vertical bonds with a magnitude of $V/\sqrt{2}$. We recall, that for a correct analysis of the critical behavior in the $GCE$, the Hamiltonian~(\ref{eq1}) has to be extended with a term $-\!\mu\hat{n}_{k}$, involving the chemical potential $\mu$, which controls the bond electron density $n_k$.
To study the critical behavior of the investigated model in both ensembles, we use the  rigorous approach based on the decoration-iteration transformation~\cite{Fisher,Rojas,Strecka2}. This procedure, described in detail in Refs.~\cite{Cenci1,Cenci2}, results to the unique relation $\Xi(T,J,J',t,V)\!=\!A^{2N}Z_{IM}(T,R)$ between the grand-canonical or canonical partition function $\Xi$ of the investigated model and the canonical partition function $Z_{IM}$ of a pure Ising model with new effective parameters $A$ and $R$. Based on their knowledge we are able to determine the critical temperature $T_c$ as a solution of the critical condition $\sinh^2(2\beta_c R)\!=\!1$, in which $\beta_c\!=\!1/k_BT_c$  and $k_B$ is a Boltzmann constant. 
\vspace*{-0.25cm}
\section{Results and discussion}
\begin{figure}[h!]
{\includegraphics[width=0.24\textwidth,height=3.4cm,trim=3.25cm 9.5cm 3.6cm 9.5cm, clip]{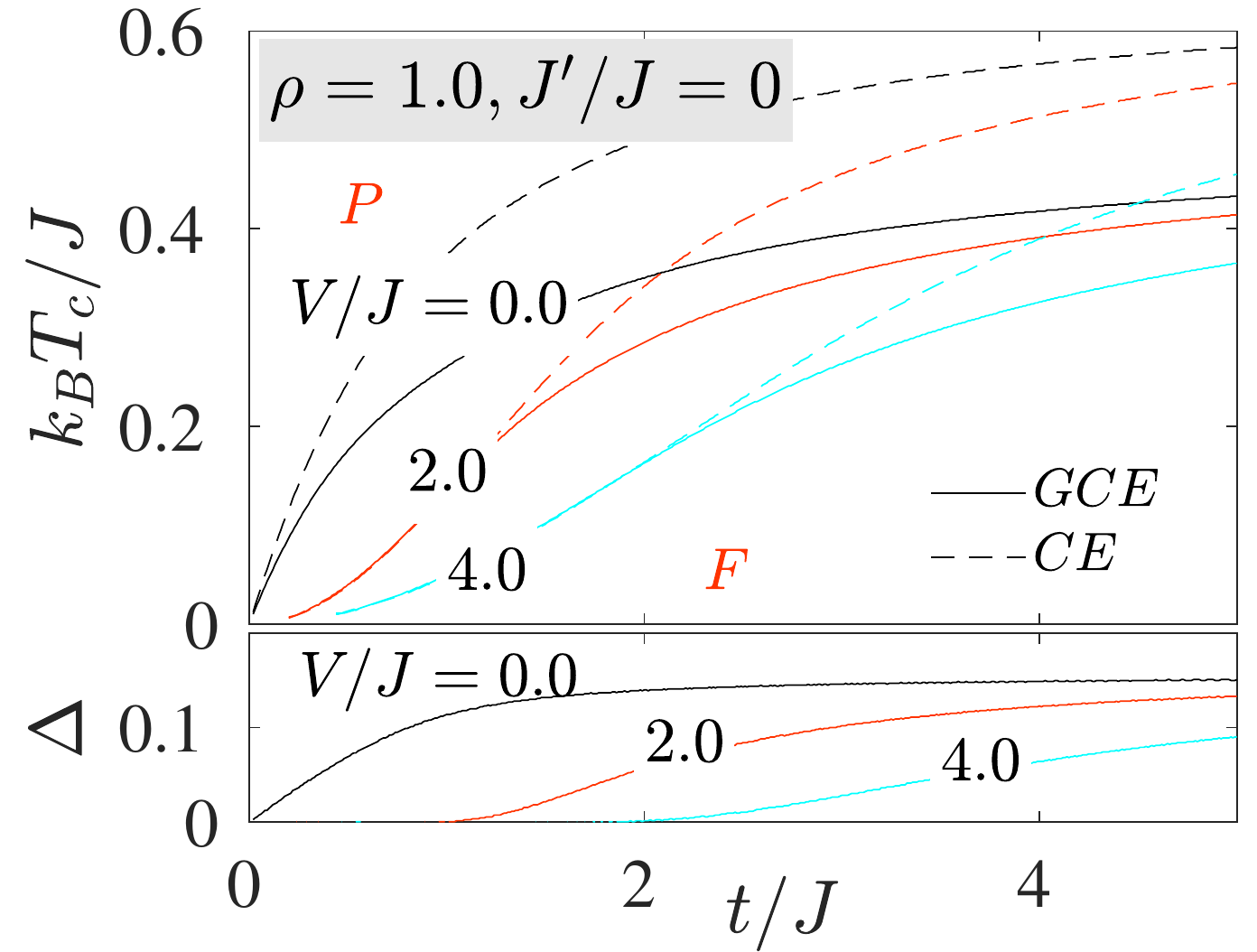}}
{\includegraphics[width=0.24\textwidth,height=3.4cm,,trim=3.25cm 9.5cm 3.6cm 9.5cm, clip]{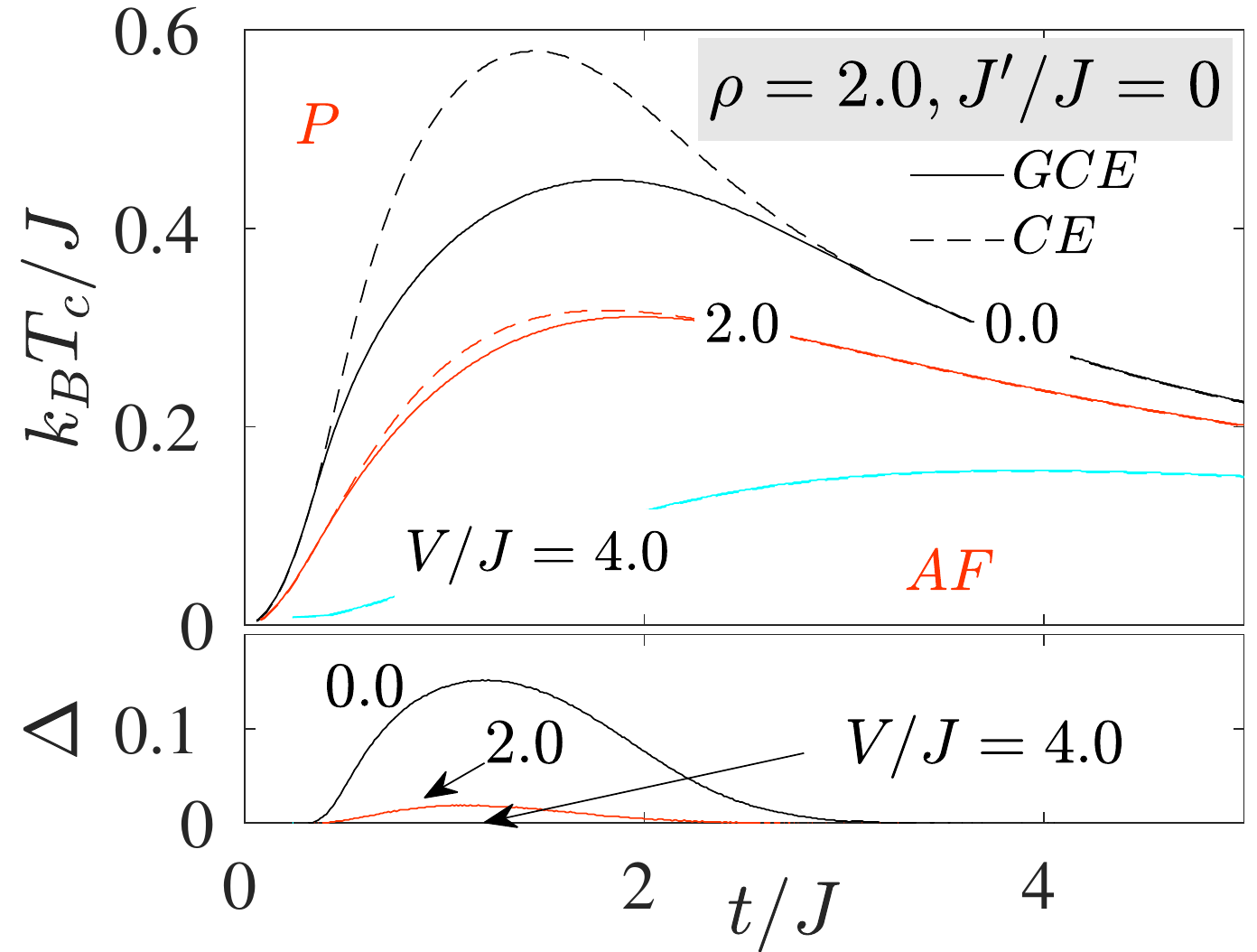}}
\includegraphics[width=0.24\textwidth,height=3.4cm,trim=3.25cm 9.5cm 3.6cm 9.5cm, clip]{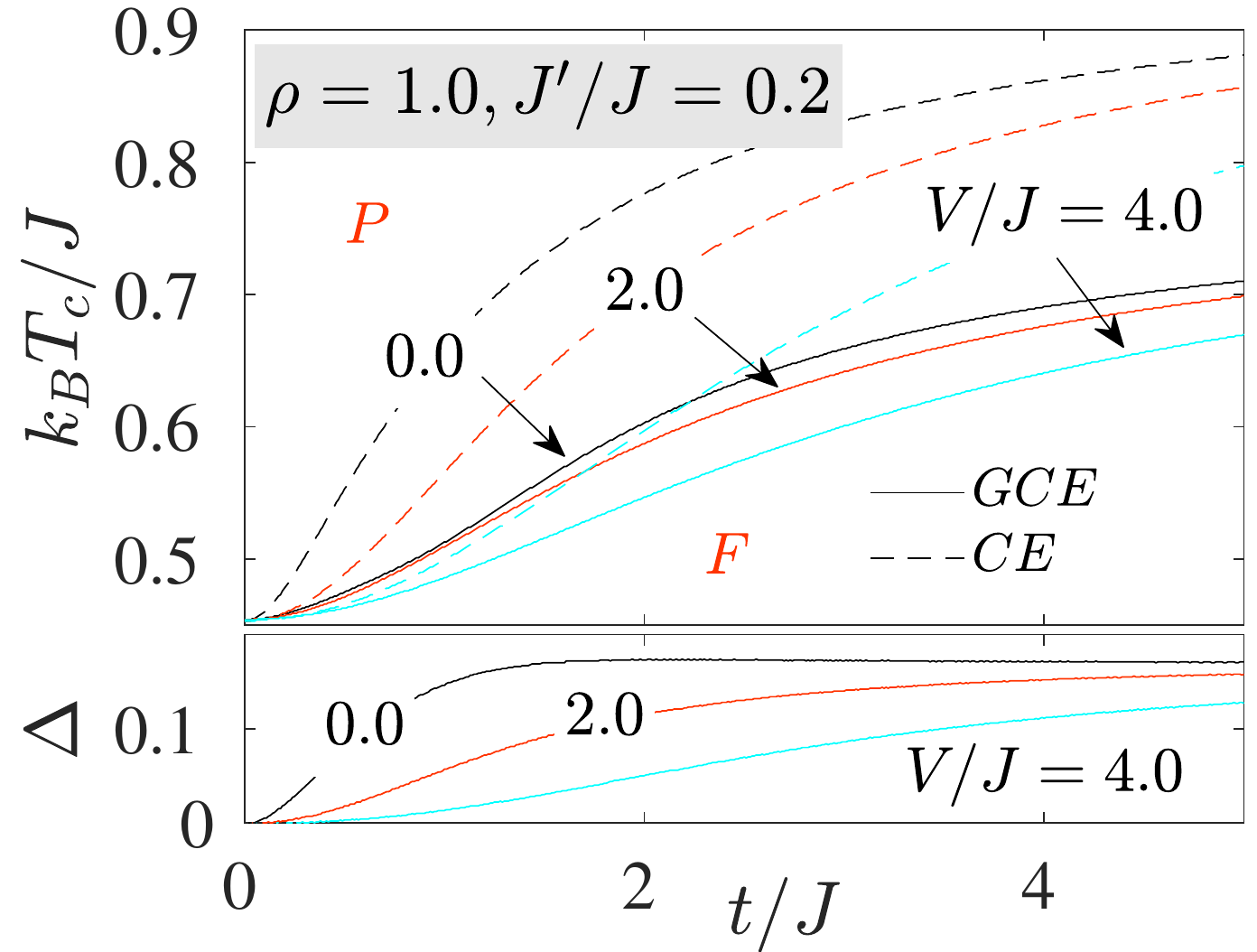}
{\includegraphics[width=0.24\textwidth,height=3.4cm,trim=3.25cm 9.5cm 3.6cm 9.5cm, clip]{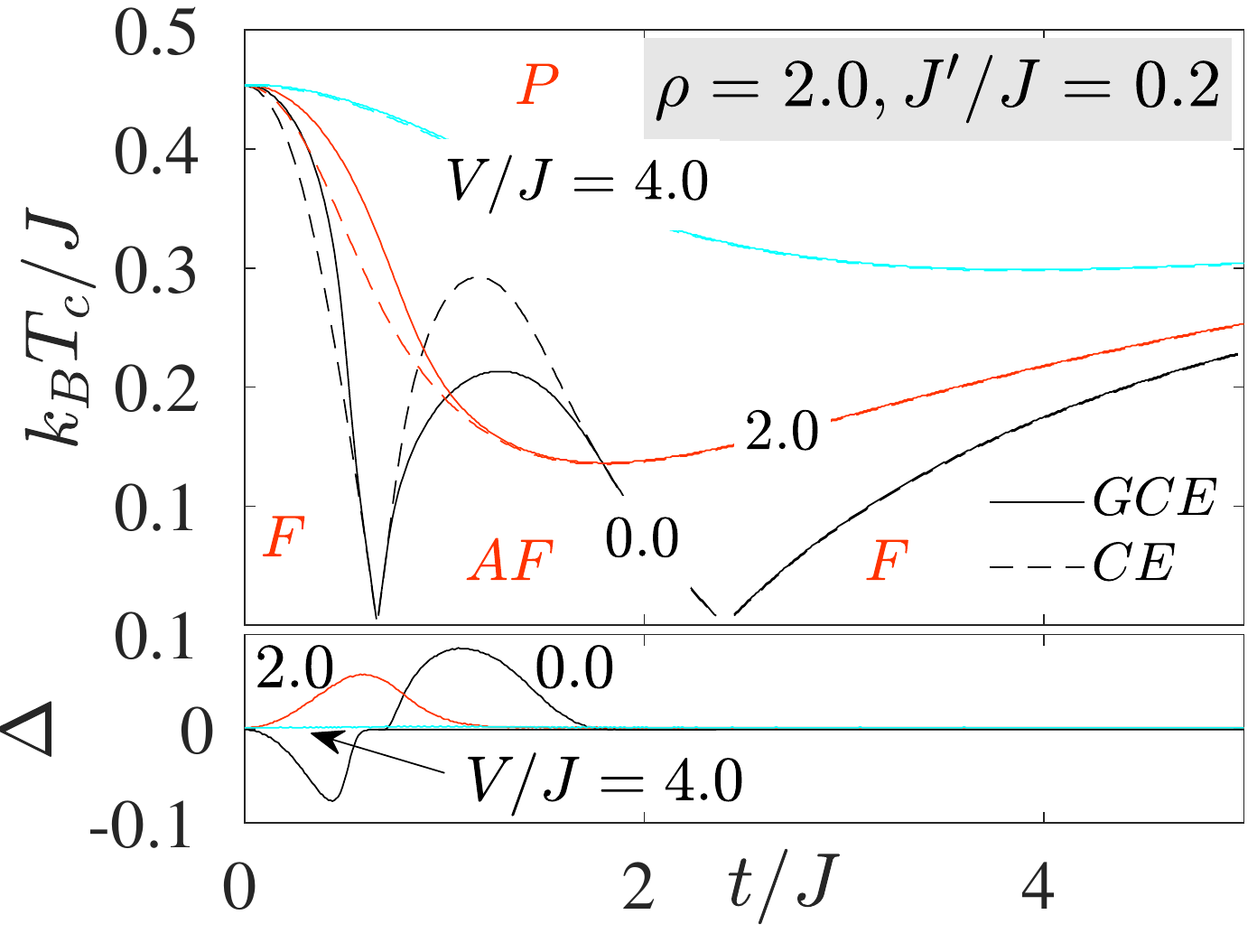}}
\includegraphics[width=0.24\textwidth,height=3.4cm,trim=3.25cm 9.5cm 3.6cm 9.5cm, clip]{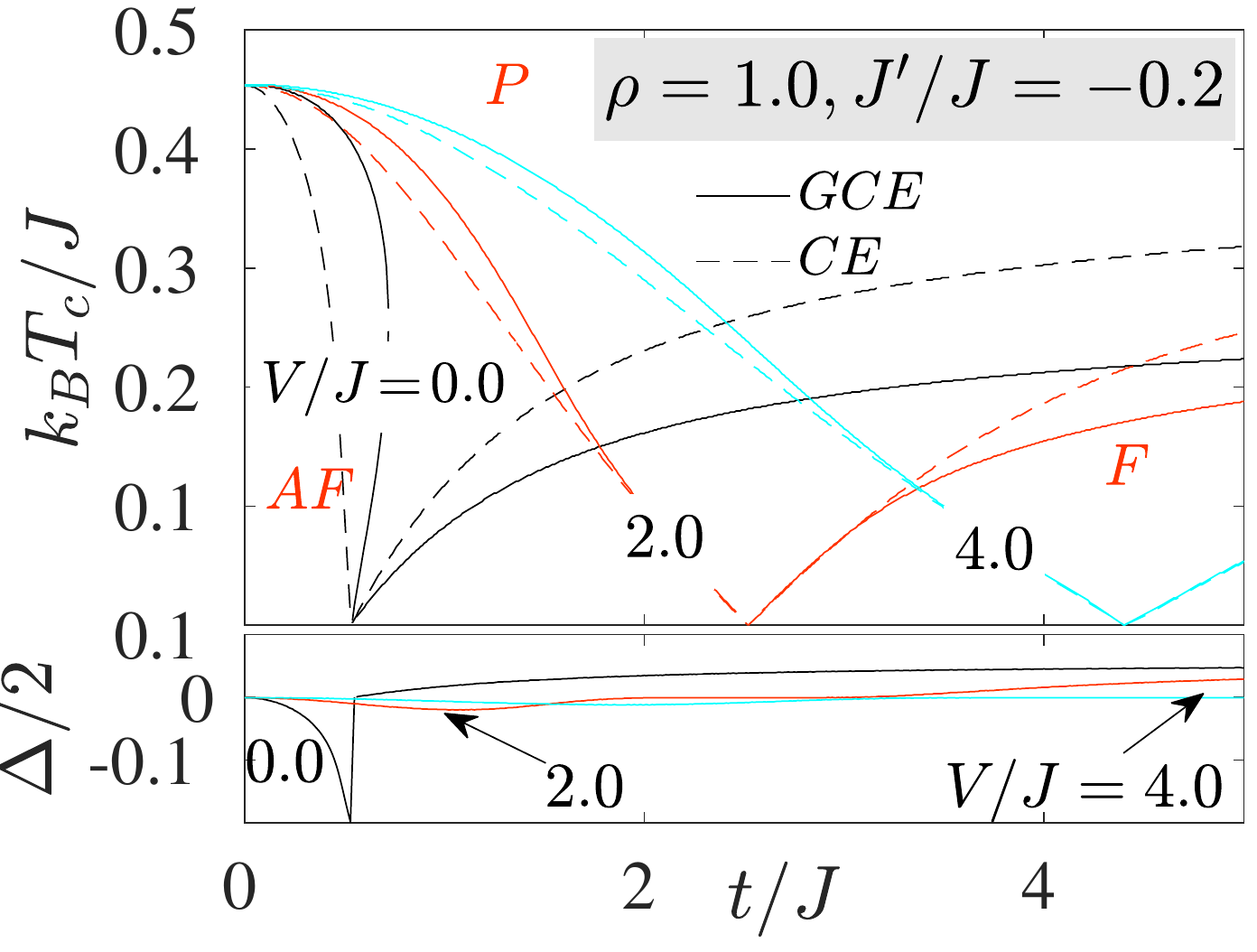}
{\includegraphics[width=0.24\textwidth,height=3.4cm,trim=3.25cm 9.5cm 3.6cm 9.5cm, clip]{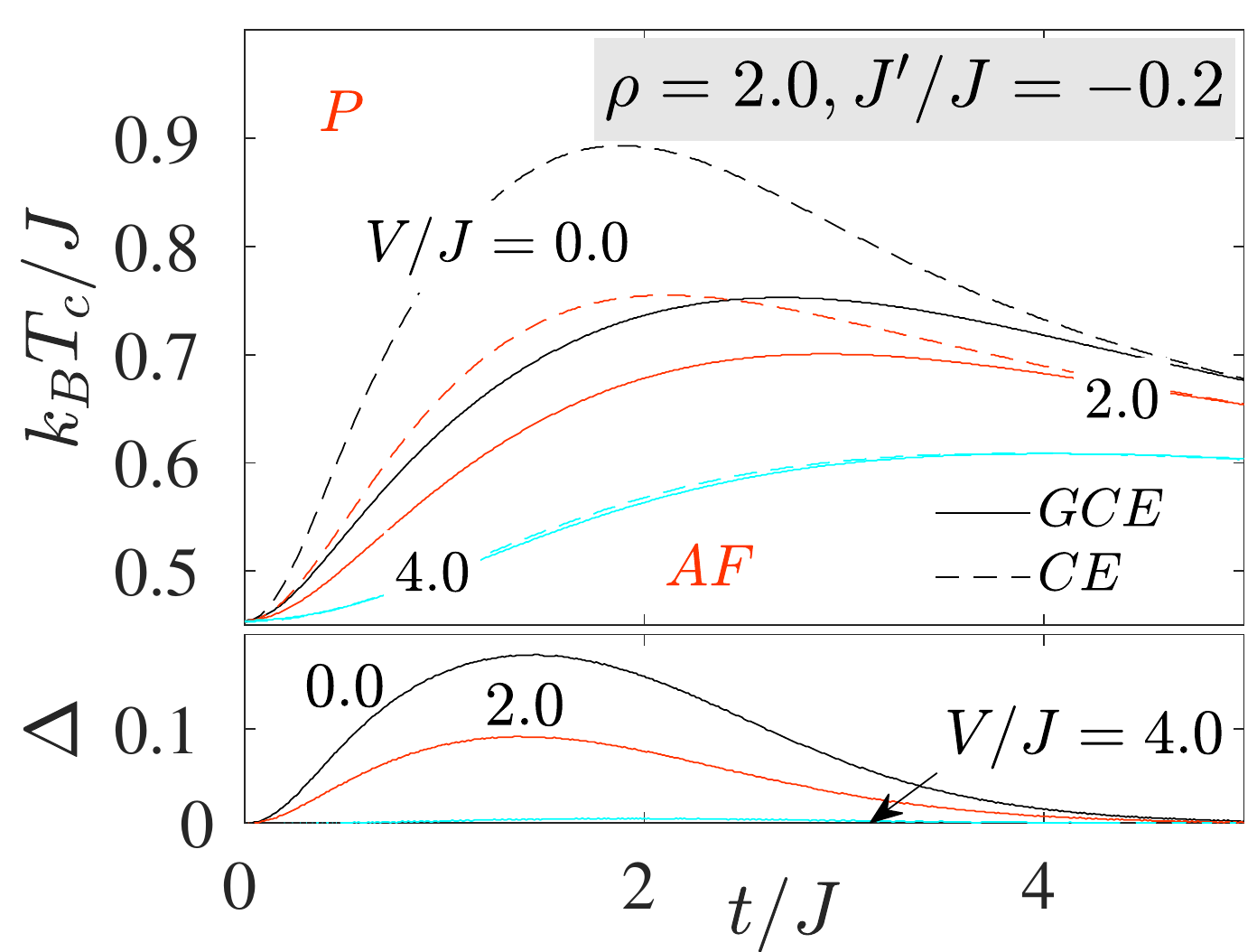}}
\caption{\small Finite temperature phase diagrams in the $k_BT_c/J\!-\!t/J$ plane for several values of model parameters. $GCE$ (solid lines) and $CE$ (dashed lines) results. Lower parts of each figures present a behavior of $\Delta$ against the $t/J$.}
\label{fig1}
\end{figure}

\vspace*{-0.25cm}
Before a detailed discussion, we note that all analyzes are performed for a ferromagnetic (F) Ising interaction $J\!>\!0$, since the antiferromagnetic (AF) one leads to the preference of the physically identical magnetic structures  in which each Ising spin is replaced with its equivalent  oriented antiparallelly to the electron spins. Two particular electron concentrations  $\rho\!=\!1$  and $\rho\!=\!2$ are assumed, where $\rho\!=\!\sum_k^{2N} n_k/2N$. The most interesting results are collected at Fig.~\ref{fig1}, where the normalized critical temperature $k_BT_c/J$ against the electron hopping $t/J$ is plotted for both electron concentrations.
The detail inspection of the $J'/J\!=\!0$ case feels out an evident ensemble inequality between critical temperatures $\Delta\!=\!k_B(T_c^{CE}\!-\!T_c^{GCE})/J$ for both investigated $\rho$. Obviously, the system with a fixed number of particles ($CE$) is a more resistant to  thermal fluctuations as its  counterpart with a variable number of particles ($GCE$), resulting to the equal or higher value of $k_BT_c/J$. The possible explanation of this interesting observation lies in the fact, that in the $GCE$, the system with a defined $\rho$ is a mixture of totally 2N independent bonds with  different eigenvectors conditioned by various number of electrons  ranging from $n_k\!=\!0$ to $n_k\!=\!2$ ($\rho\!=\!1$), or to $n_k\!=\!4$ ($\rho\!=\!2$). These different eigenvectors exhibit also a different magnetism,  F or AF one~\cite{Cenci1} and thus, the system is slightly 'disordered'. Due to this fact, the magnetic ordering of the system  in 
 the $GCE$ is destroyed easier   than  in the $CE$,  where all 2N bonds of the investigated model exhibit just the one type of magnetic ordering.  In addition, it is clear from Fig.~\ref{fig1} that the ensemble inequality $\Delta$ strongly depends on  the type of the ground-state arrangement. Whereas in the F phase the  $\Delta$ increases with increasing $t/J$ up to its saturation value at $t/J\!\to\!\infty$, in the AF one the  $\Delta$ reaches  nonzero values  exclusively at small or  intermediate  values of $t/J$.  
We suppose that the origin of this different behavior is attributable to a charge distribution over the bond upon the $t/J$ modulation resulting to  changes of an AF and non-magnetic (NM) character of respective electron dimers with $n_k\!=\!2$. It is noteworthy to mention that such electron dimers can be likewise found in a quarter filling if the  $GCE$ is taken into account. 
Moreover, it is detected that the ensemble inequality $\Delta$ can be easily reduced by switching on the $V/J$, which similarly as $t/J$ favors charge segregation instead of the homogeneous electron distribution.

The evident ensemble inequality in a critical temperature is also detected for a nonzero interaction $J'/J$, see Fig.~\ref{fig1}. Moreover, in the regime where the  $|J'/J|\!>\!0$ preserves merely the one type of ground-state magnetic structure with a selected electron concentration~\cite{Cenci2}
 (e.g., $\rho\!=\!1, J'/J\!=\!0.2$ and $\rho\!=\!2, J'/J\!=\!-0.2$) a direct coherence between Ising spins leads to the enhancement of the magnitude as well as the area of $\Delta$. The qualitative character of all previous conclusions remains unchanged.  The situation is slightly different in the regime, where the competition among all model parameters leads to the existence of  quantum magnetic phase transitions driven by the electron hopping $t/J$. In this regime the ensemble inequality $\Delta$ can reach besides positive values also  negative ones exclusively detected at $t/J\!\to\!0$. The explanation of 
an enhancement of a critical temperature at $\rho\!=\!2$ in the $GCE$ against the $CE$ may be found in a fully F character of bonds with an odd number of $n_k$ rarely distributed over the remaining double occupied  bonds with a parallel orientation of Ising spins accompanied by a superposition of electrons in AF and NM states. On the other hand, an enhancement of a critical temperature at $\rho\!=\!1$ in the $GCE$ against the $CE$ has its origin in a presence of a few AF bonds with $n_k\!=\!2$, which dilute remaining bonds with $n_k\!=\!1$ arranged in novel AF formation~\cite{Cenci2}. The  mixtured AF/NM character of the electron subsystem at AF bonds with $n_k\!=\!2$ has a strengthening effect on the stability of magnetic arrangement in contradiction to the F one observed at $n_k\!=\!1$. Similarly as in the $J'/J\!=\!0$ case,  the nonzero electrostatic potential $V/J$ suppresses the ensemble inequality in a critical temperature, however the reduction effect is slightly damped by nonzero interaction between nearest-neighbor Ising spins, see Fig.~\ref{fig1}.

Another interesting observation which directly follows from the Fig.~\ref{fig1} is an absence of the thermally driven re-entrant magnetic phase transition in the $CE$, which is a very specific feature detected on the DDSEM with a square plaquette in the $GCE$~\cite{Cenci2}. The representative situation is shown in Fig.~\ref{fig1} for  $\rho\!=\!1$, $J'/J\!=\!-0.2$ and $V/J\!=\!0$, however, an identical observation has been detected in a whole parameter space, where the thermally driven re-entrant magnetic-phase transition has been detected assuming the $GCE$. Based on this interesting observation, we suppose that the thermally driven re-entrant magnetic-phase transition observed in real materials is an intrinsic feature of materials and originates from the different number of valence electrons between two nearest-neighbor localized magnetic ions. Consequently, to bring a deeper insight into the understanding of this unconventional phenomenon, the respective theoretical analyzes should be performed using the $GCE$.
\vspace*{-0.25cm}
\section{Conclusions}
\vspace*{-0.25cm}
We have performed a comparative rigorous study of a critical behavior of the DDSEM on a square lattice in the $CE$ and $GCE$.  It has been found that there exist regions, where the critical temperature and thus the stability of magnetically ordered phases can be dramatically different in the system with a fixed ($CE$) or fluctuating ($GCE$) number of particles.  As was discussed, the crucial impact on the magnitude of $\Delta$ lies in the number of doubly occupied bonds with a charge segregation at one of two decorating sites, which can support or reduce the effect of thermal fluctuations on the stability of magnetic orderings. In addition, it was demonstrated that the intriguing thermally driven re-entrant magnetic phase transition observed in real materials can be theoretically described only by the $GCE$  and is fully absent in the $CE$ counterpart. Finally, it was shown that the ensemble inequality $\Delta$  is gradually reduced by applying the external electric field along the \textbf{[11]} crystallographic axis.
%
\\\\
This work was supported under the grant Nos. APVV-16-0186, VEGA 1/0043/16 and  ITMS 26220120047.
%


\end{document}